\documentclass[a4paper]{jpconf}
\usepackage{graphicx}
\usepackage{url}

\def\ha{H$\alpha$}

\def\HII{H\,\textsc{ii}}
\def\NII{[N\,\textsc{ii}]}
\def\OII{[O\,\textsc{ii}]}

\def\SR{$S_{\rm H\alpha}$--$r$}

\begin{document}

\title{The H$\alpha$ surface brightness -- radius plane as a diagnostic tool for photoionized nebulae}

\author{
David J  Frew$^{1,2}$, Ivan S Boji\v{c}i\'c$^{1,2}$ and Quentin A  Parker$^{1,2}$
}

\address{$^1$ Department of Physics, The University of Hong Kong, Hong Kong SAR, China}
\address{$^2$ Laboratory for Space Research, The University of Hong Kong, Hong Kong SAR, China}

\ead{djfrew@hku.hk}

\begin{abstract}
The H$\alpha$ surface brightness -- radius (\SR) relation is a robust distance indicator for planetary nebulae (PNe), further enhanced by different populations of PNe having distinct loci in \SR\ space. Other types of photoionized nebulae also plot in quite distinct regions in the \SR\ plane, allowing its use as a diagnostic tool. In particular, the nova shells and massive star ejecta (MSE) plot on relatively tight loci illustrating their evolutionary sequences.  For the MSE, there is potential to develop a distance indicator for these objects, based on their trend in \SR\ space. As high-resolution, narrowband surveys of the nearest galaxies become more commonplace, the \SR\ plane is a potentially useful diagnostic tool to help identify the various ionized nebulae in these systems.
\end{abstract}

\section{Introduction}
This paper explores the potential of the \SR\ plane as a diagnostic tool for the study of photoionized emission nebulae. As part of our work on determining distances for Galactic planetary nebulae (PNe), we developed and calibrated the \ha\ surface brightness -- radius (\SR) relation \cite{FPB16}, which has been shown to be more accurate than previous statistical distance methods, particularly for the demographically common `senile' PNe.  Its application  requires an angular size, an integrated \ha\ flux, and the reddening to the PN.  From these quantities, an intrinsic radius is calculated, which when combined with the angular size, yields the distance directly.  Rather than using the radio domain, we chose the \ha\ emission-line, firstly as it best represents the nebular ionized mass, and secondly because a number of narrowband \ha\ imaging surveys have recently become available \cite{SHASSA}\cite{WHAM}\cite{Drew05}\cite{SHS}, from which accurate integrated fluxes and/or diameters can be determined \cite{FBP13}\cite{F14a}\cite{Madsen}.  Ongoing \ha\ surveys \cite{Drew14} will continue to be useful, as part of our ongoing efforts to characterize the entire Galactic PN population \cite{P15}.

\section{The \ha\ relation for planetary nebulae}

In \cite{FPB16} we described in detail the construction of  a catalogue of \ha\ fluxes, angular diameters, and distances for both Galactic and extra-galactic PNe, to be used as primary calibrators  for the \SR\ relation.  This relation was refined from earlier versions \cite{Pierce}\cite{FPR06}\cite{FP06}\cite{FP07}\cite{F08} published previously, based on continued improvement in the input data, with consequent reduction of the uncertainties. %Comment on radio data
A range of criteria to further improve the precision of the \SR\ relation were investigated by \cite{F08}, with \cite{FPB16} eventually separating PNe into two broad groups based on spectroscopic criteria.  Optically thick PNe (with strong \OII\ and \NII\ lines) are systematically more massive than the optically-thin PNe that fall along the lower part of the PN locus.  Using sub-trends has allowed more precision in determining distances, as good as 18 per cent in the case of optically-thin PNe.    
Recently, \cite{Smith15} analysed the commonly-used SSV distance scale \cite{SSV} in some depth.  In particular, this scale has a substantial scale error at large PN radii, meaning that the distances for evolved (demographically-common) PNe are considerably underestimated, by a factor of two-or-so.  Our revised scale \cite{FPB16} is largely free of this problem, having improved on the mean \SR\ relation of \cite{F08}, which in turn had been independently validated \cite{Smith15}\cite{JSS}\cite{Ali} as the most reliable statistical distance scale to date.   
We expect our distances to remain useful even after the expected data avalanche from the Gaia satellite \cite{Gaia}, as many PN central stars are fainter than the Gaia magnitude limit, or have confusion problems in compact PNe of high surface brightness \cite{Mant}.  Of course the Gaia parallaxes will allow the refinement of our proposed sub-trends in the $S$--$r$ plane, enhancing its ability both as a diagnostic tool, and as a robust measure for the many PNe without trigonometric distances.

%%%%%%%%%%%%%%%%%%%%%%%%%%%%%%%%%%
\section{Other Emission Nebulae}

\begin{figure*}
\centering  
\includegraphics[width=15cm]{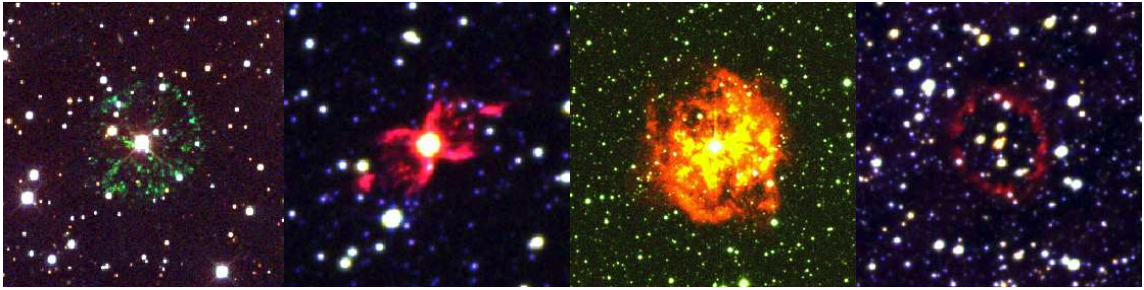}
\caption{
{\small
Color images of four `PN mimics' taken from \cite{P15}: from left, the nova shell around GK\,Per, the bipolar symbiotic outflow Hen\,2-104, and the WR ejecta M\,1-67 and PCG\,11.  This mosaic illustrates the morphological confusion between PNe and other kinds of stellar ejecta. }
}
\label{fig:mosaic}
\end{figure*}

Besides the PNe discussed in \cite{FPB16}, we are also interested in the ability of the \SR\ diagram to discriminate between bona fide PNe, transitional objects, and the zoo of PN-like nebulae and outright mimics (see Fig.\,\ref{fig:mosaic}) that have been confused with them \cite{FP10}\cite{FM10}, both in the Milky Way and in the nearest galaxies.  It is well known that some bipolar PNe have similar morphologies to the outflows around D-type symbiotic stars (SyS), and their similarities and differences have been discussed several times in the literature, e.g. \cite{CS95}\cite{C03}\cite{Kwok03}\cite{M12}.  To investigate these nebulae, we have adopted \ha\ fluxes from the literature, supplemented with our own data \cite{FBP13}\cite{F14a}\cite{P15}, while distances have been taken from the sources given in \cite{FPB16}, supplemented with a few distances from \cite{VFP15}.   Note that the locations of the SyS nebulae in the \SR\ plot are only indicative, as their fluxes and dimensions are difficult to measure precisely.  For the ejecta around luminous blue variable (LBV) and Wolf-Rayet stars \cite{Chu03}\cite{Weis}, we adopt the data directly from \cite{FPB16}, as we do for the low-mass \HII\ regions in the ISM.  We also plot a range of nova shells where we utilize the rather heterogeneous data compiled by \cite{DD2000}\cite{DDD}\cite{Gill2000}.
Finally, we add the faint bipolar nebula surrounding CK~Vul \cite{Hajduk07}\cite{Hajduk14}, the bowshock nebulae around the nova-like cataclysmic variables (CVs) BZ\,Cam \cite{Greiner}\cite{Hollis} and V341\,Ara \cite{F08}\cite{FMP06}, as well as the peculiar object Te\,11 \cite{Jacoby10}\cite{Misz16}.

%%%%%%%%%%%%%%%%%%%%%%%%%%%%%%%%%%
\section{Discussion}

\begin{figure*}
\centering  
\includegraphics[width=14.6cm]{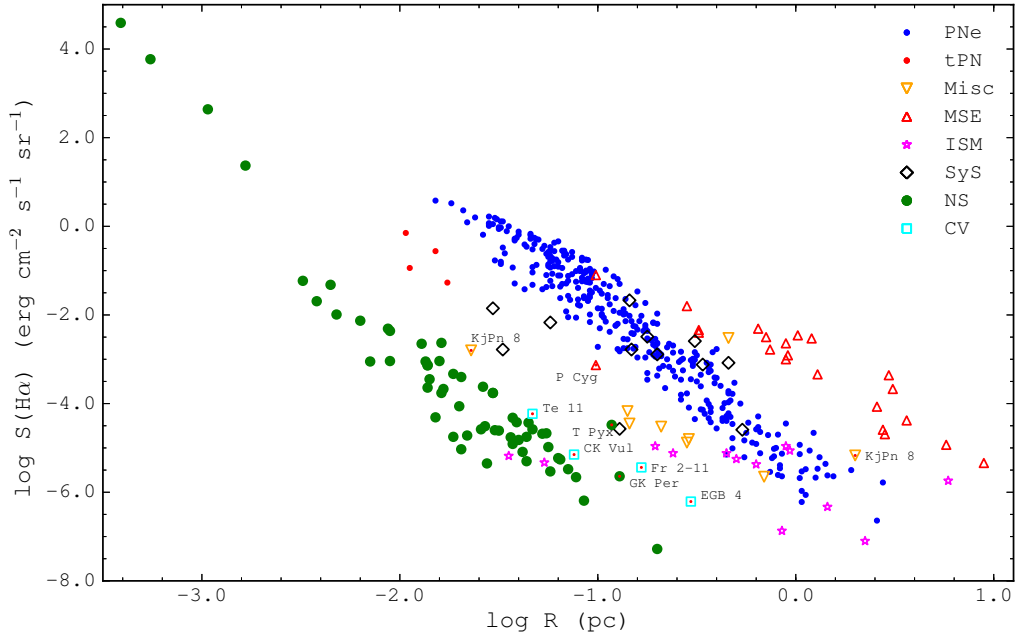}
\caption{
{\small
PNe and mimics plotted in the \SR\ plane.  Transitional PNe (tPN), miscellaneous objects (Misc), massive star ejecta (MSE), low-mass \HII\ regions in the  ISM, symbiotic (SyS) outflows, nova shells (NS), and CV nebulae have been plotted separately. Individual objects discussed in the text are labelled.}
}
\label{fig:SR_plane}
\end{figure*}

In Fig.\,\ref{fig:SR_plane}, we plot a range of different photoionized nebulae in \SR\ space. Besides the primary PN locus, some transitional PNe and miscellaneous PN-like objects described in \cite{FPB16} are also plotted.  Particularly interesting are the locations of the core and lobes of the outflow KjPn8 \cite{Lopez}, showing the large range in ejecta mass between outflow episodes.
The massive star ejecta (MSE) generally plot above the PNe, reflecting their larger ionized masses in the mean.  For the MSE, a reasonably tight evolutionary trend is seen, if we exclude the young, low-mass outflow around the historical LBV, P\,Cygni.  The points can be fit by a relation with a slope of $-2.3$, markedly shallower than the PN locus, indicating the substantial amount of circumstellar and/or interstellar material being swept up in the evolved MSE.  The data suggest an approximate distance scale can be developed for ejecta around LBVs and WN stars, at least for those shells that are not dominated by snow-plowed interstellar matter. 

The nova shells, as expected, show a steep relation (power law slope $\simeq -5$) at young ages due to adiabatic expansion, before the trend flattens at larger radii as additional circumstellar material is swept up.  The unusual (massive) shells around GK\,Per and T\,Pyx \cite{Liimets}\cite{Shara12}\cite{Shara15} are seen to be above the mean shell trend.  Note there is at least one PN hosting a classical nova \cite{Wesson}, as well a very faint bipolar nebula \cite{Shara12}\cite{Bode} surrounding the nova ejecta from GK~Per, with properties similar to a low-mass PN \cite{FPB16}.  
The CV bowshock nebulae (EGB\,4 and Fr\,2-11) are clearly seen to be of lower ionized mass than PNe, though apparently distinct to classical nova shells. Te\,11 has affinities with these nebulae, based on its nebular abundances and very low expansion velocity; while it plots close to the ordinary nova shells, its distinct morphology (i.e. a high volume filling factor) suggests a greater ionized mass and different origin, cf. \cite{Misz16}.  The bipolar nebula around CK\,Vul is seen to have a mass comparable to a nova shell, but its origin remains uncertain \cite{Hajduk14}.
As shown in \cite{FPB16} the diffuse \HII\ regions in the ISM ionized by low-mass stars are generally of low to very-low surface brightness and plot on and around the PN locus at medium to large radii.   What is somewhat unexpected is the substantial overlap in phase space between the symbiotic outflows and PNe, with some evolved SyS nebulae being as massive as typical PNe.  Along with the emission-line diagnostic diagrams we have developed \cite{FP10}\cite{Sabin}\cite{F14b}, the \SR\ plane will also be a useful adjunct to help identify ionized nebulae in and beyond the Milky Way, particularly with the advent of deep hydrogen-line surveys using the next generation of telescopes.  A more detailed discussion will be published elsewhere.

%\subsection{Acknowledgments}

\end{document}